# Latin squares as a model of perfect mixing of high-entropy alloys

P.S. Grinchuk
A.V. Luikov Heat and Mass Transfer Institute of the National Academy of Sciences of Belarus.
gps@hmti.ac.by. ORCID: 0000-0002-4124-3186

The structure of high-entropy alloys, which include 5 or more components in equimolar proportions, is proposed to be described in the mathematical language of Latin squares. This model is fundamentally discrete and allows for a quantitative description of ideal mixing, in which any given alloy element is guaranteed not to have an identical element as its nearest neighbor. For the two-dimensional case, the model possesses four axes of translational symmetry, which makes it possible to tile the plane with a single Latin square and, based on the properties of this square, describe the properties of macroscopic material samples. The model makes it possible to make quantitative estimates of the suitability of a particular atomic structure (configuration) for implementation in a perfectly mixed high-entropy alloy with maximum mixing entropy. A condition has been obtained for acceptable fluctuations in the concentration of components at which the thermodynamic stability of the alloy is maintained. The possibility of identifying superstructures in high-entropy alloys based on group theory, as applied to the description and construction of Latin squares, is discussed. Based on the proposed model, the probability of forming a multicomponent alloy with ideal mixing is estimated. For a five-component alloy, this probability is no more than $P(5) \approx 2.6 \times 10^{-10}$.

Keywords: high-entropy alloys; Latin square; Sudoku puzzle; perfect mixing; superstructure

The concept of high-entropy alloys as a new generation of functional materials emerged more than 20 years ago [1]. Such a material is an alloy consisting of five or more components in equiatomic proportions. A number of new properties inherent in such materials are due to the increased number of components. Therefore, it is expected that the properties of these alloys will surpass those of traditional alloys. Over the years, thousands of publications and dozens of high-quality reviews have appeared on the topic of high-entropy alloys and ceramics [2-5]. One of the earliest works defines high-entropy alloys as "*consisting of five or more main elements in equimolar ratios*" [1]. However, in the same paper, the authors immediately broaden the definition, relaxing it to the condition that the alloy must contain "*main elements with a concentration of each element ranging from 35 to 5 atomic %*". Thus, ever since the pioneering works, it has been generally accepted that high-entropy alloys do not necessarily have to be equiatomic. In our view, this is a debatable issue, since it is precisely due to high mixing entropy that the unique properties of these alloys are achieved. And the maximum of this entropy, as we will demonstrate again below, is reached at equimolar concentrations. On the one hand, deviating from equal concentrations expands the number of possible new alloys. But what are the permissible concentration limits, and will such alloys differ significantly from traditional alloys based on one or two components, doped with additives of other elements? Therefore, the focus of this work is on alloys with element concentrations close to equiatomic. This work is devoted to the problem of describing the uniformity of mixing of several discrete components. Note that the problem of uniform mixing under consideration goes beyond the scope of classical problems in physics (e.g., gas mixing [6]) and the description of high-entropy alloys. It is also relevant to problems involving the combustion of solid composite fuels [7] and the mixing of multicomponent powders with similar particle sizes [8-10]. To ensure a more comprehensive presentation, we will briefly review the main characteristics of high-entropy alloys.

In fact, the concept of high-entropy alloys gradually emerged from studies of bulk amorphous alloys, or metallic glasses [11, 12]. Such glasses may consist of a large number of components. In some cases, the concentrations of these components may be comparable. After crystallization, due to the high cooling rate, such alloys form an amorphous phase with a disordered atomic arrangement. When such a metallic glass is reheated, it can transition to a crystalline state, sometimes decomposing into several phases. Upon heating, this phase devitrifies - that is, it transitions to a crystalline state, sometimes decomposing into several phases in the process. One of the distinguishing features of high-entropy alloys is that, upon cooling, they form simple crystalline structures corresponding to body-centered cubic, face-centered cubic, or hexahedral close-packed lattices. It is important to note that atoms of different types are arranged randomly at the sites of these crystal lattices. This means that high-entropy alloys are disordered substitution solid solutions. The disordered arrangement of all atoms at the lattice sites leads to an increase in the material's configurational entropy. This fact served as the inspiration for the name of this class of materials.

Perhaps one of the most complex and challenging issues in the creation of multicomponent alloys is the difference in the sizes of the atoms of the individual components. This difference leads to significant distortion of the alloy's crystal lattice. It is believed that the structure of high-entropy alloys occupies an intermediate position between stable crystalline phases with a low concentration of defects and metastable metallic glasses with an amorphous structure, where long-range order is absent [4]. This structural feature of high-entropy alloys results in low diffusion coefficients, corrosion resistance, low thermal conductivity, and reduced electrical conductivity. In addition to other applications that are extensively discussed in the scientific literature, this combination makes the search for new candidates among high-entropy alloys for use as resistive materials in high-temperature electric heaters a promising avenue [13].

It is well known that certain constraints exist on the ratio of atomic sizes, their valence, and electronegativity for the formation of a substitutional solid solution. We will not dwell on this point in detail. We will merely note that these conditions are expressed by the Yum-Rosery rules [4,14]. For a multicomponent system, a simple solid solution forms only if the sizes of the atoms differ by no more than ~4% in each pair [4]. This is a fairly strict condition that significantly reduces the set of possible combinations. In addition to size compatibility, there are three other conditions. The second relates to the electronegativity of the atoms; they must be similar. If they differ significantly, then instead of solid solutions, chemical compounds in the form of intermetallics are formed. The third condition concerns the identical or similar valence of the atoms in the system. Finally, the fourth condition requires that the elementary metallic components have identical crystal structures.

An additional condition for the formation of high-entropy alloys was the enthalpy condition, which was derived empirically based on an analysis of existing alloys. The limiting values of the mixing enthalpy of elements were approximately determined; within these limits, a disordered solid solution is formed [4]: $-15\ \ldots -11.6 \leq \Delta H_{mix} \leq\ 3.2\ \ldots\ 5$ kJ/mol. Note that, in physical terms, this is the change in enthalpy upon mixing several elements, when a disordered solid solution forms after mixing.

The set of rules described does not follow directly from any physical laws or principles but is, rather, empirical in nature. Nevertheless, they are useful as criteria for the initial selection of alloy compositions.

An important consequence of the alloy's multicomponent nature - and even a slight difference in atomic sizes - is the impossibility of forming a regular crystal lattice from a macroscopic number of atoms. The requirement for an equal number (or identical set) of atoms in the alloy necessitates variations in the composition and sizes of the basic unit cells from which it is composed. Therefore, the structure of an equiatomic high-entropy alloy must be either a completely disordered substitutional solid solution or a solid solution with short-range order. It is important to note that such solutions can become ordered during annealing - sometimes over a sufficiently long period - due to the slow diffusion of atoms in a disordered environment.

Eventually, all of the rules and limitations described lead to the formation of a unique composition of high-entropy alloys. These characteristics can be summarized as several key effects: high mixing entropy, the crystal lattice distortion effect, the slowed diffusion effect, and the "cocktail effect" [4,15].

A high-entropy alloy, like any other, will be in a stable state if its Gibbs potential (Gibbs energy), G = H - TS, is minimized. Here, H is enthalpy, T is absolute temperature, and S is entropy. From the minimization condition for an isobaric-isothermal system (ΔT = 0), the well-known condition follows:

$$\Delta G = \Delta H - T\Delta S \; > 0 \qquad (1)$$

This leads to an important condition for the system's entropy. Only those structural changes are thermodynamically allowed in which the decrease in entropy S as the system (alloy) transitions to a more ordered state at a given fixed temperature T exceeds the decrease in enthalpy H due to the exothermic effect of the formation of the ordered phase:

$$\Delta S \; > \; -\frac{\Delta H}{T} \qquad (2)$$

To fulfill this condition, the entropy of the system must be large enough. A number of properties of high-entropy alloys allow us to estimate it. To do so, we must use the Boltzmann equation, which expresses entropy in terms of the number Ω of microstates (configurations) that realize it. Configurational entropy is the part of a system's entropy that is associated with the number of possible ways in which particles are distributed in space or with their orientational arrangement. It accounts for the variety of spatial or orientational arrangements of atoms, molecules, or other structural units within the system.

The basic formula for calculating configurational entropy is the Boltzmann equation:

$$S_{conf} = k_B \; ln \, \Omega \qquad (3)$$

where $k_B$ is the Boltzmann constant. More generally, if a system can be in n states with probabilities $P_n$, then the ideal configuration entropy of mixing is given by the Gibbs entropy formula:

$$S_{conf} = -R \; \sum_{n=1}^{N} P_n \; ln(P_n). \qquad (4)$$

In the limit of perfect disorder (when all probabilities are equal), the Gibbs formula reduces to the Boltzmann formula. A statistical analysis of a mixture of N components with concentrations $c_i$ (i=1..N) yields probabilities $P_i \; = \; c_i$. In the case of a high-entropy alloy, all concentrations are, by definition, equal: $c_1 = .. = c_N = \frac{1}{N}$. Then we can write:

$$S_{conf} \; = - \; R \sum_{i=1}^{N} \frac{1}{N} ln \frac{1}{N} = R \; ln \; N \qquad (5)$$

In the transition from a disordered solid solution to a fully ordered crystalline phase, only the configurational entropy of the system changes (tending toward zero as the number of configurations that realize the state decreases), while the other components of entropy remain unchanged. Therefore, when analyzing the stability of high-entropy alloys, a change in entropy is equivalent to a change in configurational entropy. The condition for the stability of an alloy is then expressed as:

$$ln \; N \; > \; -\frac{\Delta H}{RT} \qquad (6)$$

This equation allows us to estimate the change in the system's enthalpy during the transformation of a disordered solid solution into a fully ordered one, depending on the number of components.

Condition (6) also serves as the basis for the classification of multicomponent alloys. Thus, traditional alloys with N ≤ 2, for which $\Delta S_{conf} \leq 0.69R$, are called low-entropy alloys; alloys with $0.69R < \Delta S_{conf} < 1.61R$ (N = 3–4) are called medium-entropy alloys; alloys with $\Delta S_{conf} \geq 1.61R$ (N ≥ 5) are classified as high-entropy [15]. It is important to note that equations (5) and (6) hold for ideal solutions, whereas for regular solutions they are only approximate. Most solid solutions are irregular. Therefore, the stable compositions of high-entropy alloys may differ from equimolar ones [15]. This is one of the central issues of interest in the search for new materials with improved properties. Thus, high-entropy alloys can be found not only among equimolar compositions but also among compositions with slight deviations in concentration from the equimolar state. Let us consider how the entropy of the system changes in this case.

Suppose we have a system with small fluctuations in concentration, such that $P_{n,*} = P_n - \Delta_n$, where $P_n = \frac{1}{N}$ is the probability of finding the $n^{th}$ type of atom in a random cell. The following condition holds:

$$\sum_{n=1}^{N} \Delta_n = 0. \tag{7}$$

Then

$$\Delta S_{conf,*} = -R \sum_{n=1}^{N} P_{n,*}\, ln(P_{n,*}) = -R \sum_{n=1}^{N} (P_n - \Delta_n)\, ln(P_n - \Delta_n) \,.$$

$$(P_n - \Delta_n)\, ln(P_n - \Delta_n) \approx P_n ln(P_n) + \Delta_n P_n ln(P_n) - \Delta_n P_n \,.$$

$$\Delta S_{conf,*} \approx -R \left\{ \sum_{n=1}^{N} P_n ln(P_n) + \sum_{n=1}^{N} \Delta_n P_n ln(P_n) - \sum_{n=1}^{N} \Delta_n P_n \right\}.$$

$$\Delta S_{conf,*} \approx \Delta S_{conf} - R \left\{ \sum_{n=1}^{N} \Delta_n P_n ln(P_n) - \sum_{n=1}^{N} \Delta_n P_n \right\}.$$

Thus, by condition (7), we have:

$$\sum_{n=1}^{N} \Delta_n P_n ln(P_n) = \sum_{n=1}^{N} \Delta_n \frac{1}{N} ln(\frac{1}{N}) = -\frac{ln(N)}{N} \sum_{n=1}^{N} \Delta_n = 0, \tag{8}$$

$$\sum_{n=1}^{N} \Delta_n P_n = \sum_{n=1}^{N} \Delta_n \frac{1}{N} = \frac{1}{N} \sum_{n=1}^{N} \Delta_n = 0. \tag{9}$$

$$\Delta S_{conf,*} \approx \Delta S_{conf}. \tag{10}$$

The linear term vanishes. This is a sign that the function has reached a maximum and, consequently, that entropy has reached a maximum for an equimolar mixture. Let's find the quadratic term.

$$ln(P_n - \Delta_n) = ln\left\{P_n(1 - \frac{\Delta_n}{P_n})\right\} = lnP_n + ln\left\{1 - \frac{\Delta_n}{P_n}\right\} \approx lnP_n - \frac{\Delta_n}{P_n} + \frac{1}{2}\left(\frac{\Delta_n}{P_n}\right)^2 ,$$

$$(P_n - \Delta_n)\, ln(P_n - \Delta_n) \approx P_n lnP_n - \Delta_n + \frac{(\Delta_n)^2}{2P_n} - \Delta_n\left\{lnP_n - \frac{\Delta_n}{P_n} + \frac{1}{2}\left(\frac{\Delta_n}{P_n}\right)^2\right\},$$

$$\Delta_n\left\{lnP_n - \frac{\Delta_n}{P_n} + \frac{1}{2}\left(\frac{\Delta_n}{P_n}\right)^2\right\} \approx \Delta_n lnP_n - \frac{(\Delta_n)^2}{P_n} + \frac{(\Delta_n)^3}{2}\left(\frac{1}{P_n}\right)^2.$$

Up to a quadratic term, and taking into account Eqs. (8) and (9), we obtain the following for the change in configuration entropy:

$$\Delta S_{conf,*} \approx -R\left\{\sum_{n=1}^{N} P_n ln(P_n) - \sum_{n=1}^{N}\frac{(\Delta_n)^2}{2P_n}\right\} \quad (11)$$

Let's assume that all deviations in concentration are approximately equal:

$$|\Delta_n| = |\Delta c_n| = \frac{\Delta c_*}{N}. \quad (12)$$

Then

$$\Delta S_{conf,*} \approx \Delta S_{conf} + R\frac{N}{2}\left\{\sum_{n=1}^{N}\left(\frac{\Delta_*}{N}\right)^2\right\}.$$

$$\Delta S_{conf,*} \approx \Delta S_{conf} + R\frac{(\Delta c_*)^2}{2}. \quad (13)$$

Thus, thermodynamics allows for fluctuations in the concentrations of the components (deviations from equimolarity) that satisfy condition (14):

$$\Delta S \approx \Delta S_{conf} + R\frac{(\Delta c_*)^2}{2} > -\frac{\Delta H}{T},$$

$$\Delta S_{conf} > -\frac{\Delta H}{T} - R\frac{(\Delta c_*)^2}{2},$$

$$ln\, N > -\frac{\Delta H}{RT} - \frac{(\Delta c_*)^2}{2}. \quad (14)$$

Formally, condition (14) could also imply a significant deviation in the concentration of a single component while maintaining equimolar concentrations for all others.

Let’s now move on directly to the main model of this work. A key issue in the theory of high-entropy alloys is the lack of any advanced models to describe their structure. Most studies involve first-principles molecular dynamics simulations [16-18]. Such modeling, even for a specifically selected structure, is in itself extremely computationally expensive. If, in addition, one must choose from a large set of combinations and configurations of elements, the extreme length of the simulation completely negates its significance. A method is needed to assess the suitability of a given atomic structure (configuration) for implementation in a high-entropy alloy.

We propose using Latin squares (and, in the future, Latin cubes) as a model for the structure of high-entropy alloys. For clarity, we will consider two-dimensional structures. All conclusions and theoretical constructions will, however, be generalizable to the three-dimensional case. A Latin square is an N × N matrix filled with N elements from a set M, each of which appears exactly N times in the set. Furthermore, in each column and each row of the matrix, each of the N elements appears only once. The elements of the set can be numbers from 1 to N or, for example, Latin letters or symbols representing various atoms. From here on, a straightforward analogy applies. The N elements in an alloy must be arranged in the cells of the matrix. A one-dimensional arrangement of the elements does not adequately describe their spatial configuration. However, arranging them according to the Latin square rule can serve as a discrete model of the structure. The most important property of such a model is that arranging the elements according to the Latin square principle guarantees perfect mixing of all elements and maximum configurational entropy according to Equation (10). In this context, the diagonal arrangement of identical elements is not considered an adjacent position. An important point is the rejection of a continuous approach to describing mixing, where, for example, the probability density of finding a specific element at a point in space is considered and the concept of interpenetrating continua is allowed. The proposed approach is fundamentally discrete. Examples of Latin squares are shown in Figure 1.

| | | |
|---|---|---|
| 2 | 3 | 1 |
| 1 | 2 | 3 |
| 3 | 1 | 2 |

| | | | |
|---|---|---|---|
| 2 | 1 | 3 | 4 |
| 4 | 3 | 1 | 2 |
| 3 | 2 | 4 | 1 |
| 1 | 4 | 2 | 3 |

| | | | | |
|---|---|---|---|---|
| 4 | 2 | 5 | 3 | 1 |
| 1 | 4 | 2 | 5 | 3 |
| 5 | 3 | 1 | 4 | 2 |
| 2 | 5 | 3 | 1 | 4 |
| 3 | 1 | 4 | 2 | 5 |

| | | | | | |
|---|---|---|---|---|---|
| 4 | 6 | 5 | 2 | 3 | 1 |
| 5 | 1 | 6 | 3 | 4 | 2 |
| 1 | 3 | 2 | 5 | 6 | 4 |
| 6 | 2 | 1 | 4 | 5 | 3 |
| 3 | 5 | 4 | 1 | 2 | 6 |
| 2 | 4 | 3 | 6 | 1 | 5 |

| | | | | | | |
|---|---|---|---|---|---|---|
| 6 | 4 | 2 | 7 | 3 | 5 | 1 |
| 7 | 5 | 3 | 1 | 4 | 6 | 2 |
| 2 | 7 | 5 | 3 | 6 | 1 | 4 |
| 3 | 1 | 6 | 4 | 7 | 2 | 5 |
| 4 | 2 | 7 | 5 | 1 | 3 | 6 |
| 5 | 3 | 1 | 6 | 2 | 4 | 7 |
| 1 | 6 | 4 | 2 | 5 | 7 | 3 |

| | | | | | | | |
|---|---|---|---|---|---|---|---|
| 6 | 3 | 4 | 2 | 5 | 1 | 8 | 7 |
| 7 | 4 | 5 | 3 | 6 | 2 | 1 | 8 |
| 5 | 2 | 3 | 1 | 4 | 8 | 7 | 6 |
| 2 | 7 | 8 | 6 | 1 | 5 | 4 | 3 |
| 8 | 5 | 6 | 4 | 7 | 3 | 2 | 1 |
| 4 | 1 | 2 | 8 | 3 | 7 | 6 | 5 |
| 3 | 8 | 1 | 7 | 2 | 6 | 5 | 4 |
| 1 | 6 | 7 | 5 | 8 | 4 | 3 | 2 |

Figure 1. Examples of Latin squares of orders 3 through 8.

What does this model offer? First, it provides an accurate description of a state of ideal mixing - or even the set of all possible states of ideal mixing. This is important. It is analogous to the ideal gas model in gas physics. Second, it is easy to verify that this model possesses four axes of translational symmetry - translation of the entire N × N cell by the side length of the square (distance N) in the horizontal and vertical directions, and translation along either of the two diagonals. In other words, if the entire plane is tiled with a Latin square, then none of the elements will have neighbors of its own kind in any of the four adjacent cells - either vertically or horizontally. The entire system as a whole, constructed from a Latin square, will possess the property of ideal mixing. Moreover, if, as discussed above, the system loses the property of ideal mixing and the existence of groups or clusters of 2-3 or more identical atoms is allowed, then this model can easily be generalized to that case. For example, in this situation, in addition to all Latin squares, configurations in which the number 3 appears in any two adjacent cells are also permitted. If any combination of atoms has a thermodynamic advantage, then the Latin square model works well in this case as well. From the entire set, we retain only those that contain such a group in any configuration. Figure 2 shows an example of a square of order 5 with the "preferred" group "123" highlighted in various spatial configurations.

An important consequence of adopting this model is that the mathematical theory of Latin squares is well-established [19, 20]. These mathematical structures have been studied for over 250 years, dating back to the time of Leonard Euler. All of these findings can be applied to high-entropy alloys. It is difficult to obtain a direct, striking result in this essentially conceptual work, which outlines the main idea. However, it is possible to identify several important areas of research.

The review [4] noted that superstructures may exist in high-entropy alloys, which could be emerge in X-ray diffraction patterns. Why might such superstructures be important? It can be assumed that alloys with nontrivial middle- or long-range order (superstructure) will also possess a nontrivial and distinctive set of physical properties. Detecting such superstructures using X-ray diffraction methods presents significant methodological challenges [4]. For example, when studying the order in a four-component FeCoCrNi alloy, it was necessary to use X-ray anomalous scattering and neutron diffraction methods, since conventional CuKα diffraction proved unsuitable for this purpose [21]. The formation of superstructures can be a slow process, governed by the rules and limitations listed above. However, the translational symmetry of Latin squares mentioned above provides a tool for searching for superstructures. There are additional reasons for this.

| | | | | |
|---|---|---|---|---|
| 3 | 2 | 1 | 5 | 4 |
| 5 | 4 | 3 | 2 | 1 |
| 2 | 1 | 5 | 4 | 3 |
| 4 | 3 | 2 | 1 | 5 |
| 1 | 5 | 4 | 3 | 2 |
| 5 | 1 | 4 | 2 | 3 |
| 1 | 2 | 5 | 3 | 4 |
| 2 | 3 | 1 | 4 | 5 |
| 4 | 5 | 3 | 1 | 2 |
| 3 | 4 | 2 | 5 | 1 |
| 4 | 3 | 1 | 5 | 2 |
| 3 | 2 | 5 | 4 | 1 |
| 2 | 1 | 4 | 3 | 5 |
| 1 | 5 | 3 | 2 | 4 |
| 5 | 4 | 2 | 1 | 3 |
| 3 | 4 | 2 | 5 | 1 |
| 2 | 3 | 1 | 4 | 5 |
| 1 | 2 | 5 | 3 | 4 |
| 5 | 1 | 4 | 2 | 3 |
| 4 | 5 | 3 | 1 | 2 |
| 5 | 3 | 4 | 2 | 1 |
| 4 | 2 | 3 | 1 | 5 |
| 1 | 4 | 5 | 3 | 2 |
| 2 | 5 | 1 | 4 | 3 |
| 3 | 1 | 2 | 5 | 4 |
| 3 | 1 | 5 | 2 | 4 |
| 1 | 4 | 3 | 5 | 2 |
| 5 | 3 | 2 | 4 | 1 |
| 4 | 2 | 1 | 3 | 5 |
| 2 | 5 | 4 | 1 | 3 |

Figure 2. A Latin square of order 5 with the group 123 in various positions.

The first clue in this series is the application of group theory to the description of the structure of Latin squares. Thus, as early as 1922, in [22], methods of group theory were applied to the construction of Latin squares. A method was proposed for constructing Latin squares of order $N_1 \times N_2$ from Latin squares of orders $N_1$ and $N_2$. In our case, this means, for example, that from the structure of low-entropy or medium-entropy alloys, special cases of a higher-order structure - high-entropy alloys - can be constructed (Figure 3). This provides a direct guideline for the method of searching for superstructures in high-entropy alloys. From this follows an interesting conjecture, which, of course, requires rigorous proof. If the number of components in a high-entropy alloy is a prime number, then it is precisely in such an alloy that the discovery of a superstructure is less likely.

| | | | |
|---|---|---|---|
| 1 | 2 | 3 | 4 |
| 2 | 1 | 4 | 3 |
| 4 | 3 | 2 | 1 |
| 3 | 4 | 1 | 2 |

*a*)

| | | | | | |
|---|---|---|---|---|---|
| 1 | 3 | 2 | 4 | 6 | 5 |
| 3 | 2 | 1 | 6 | 5 | 4 |
| 2 | 1 | 3 | 5 | 4 | 6 |
| 6 | 5 | 4 | 3 | 2 | 1 |
| 5 | 4 | 6 | 2 | 1 | 3 |
| 4 | 6 | 5 | 1 | 3 | 2 |

*b*)

Figure 3. An example of a decomposed Latin square of order 4, composed of Latin squares of order 2 (a), and a decomposed Latin square of order 6, composed of Latin squares of order 3 (b).

Another area of search for superstructures is large Latin squares, obtained with additional constraints on their structure. A classic example of this is Sudoku puzzle. This is a Latin square of the order of 9 with the additional restriction that no number can appear more than once in any of the 3x3 subquads [24] (Fig. 4). In our view, Sudoku is an analogue and reflection of the method for constructing large Latin squares proposed in [22], in which a 9×9 Latin square is constructed from 9 reduced 3×3 squares (which are not Latin squares), within which the principle of perfect mixing is also realized. This is also a manifestation of the system's hidden symmetry. Surprisingly, similar methods based on solving Sudoku puzzles are used to accelerate the search for new protein structures [25].

| | | | | | | | | |
|---|---|---|---|---|---|---|---|---|
| 1 | 6 | 5 | 3 | 9 | 4 | 8 | 2 | 7 |
| 3 | 8 | 9 | 7 | 5 | 2 | 1 | 6 | 9 |
| 7 | 2 | 4 | 8 | 1 | 6 | 9 | 3 | 5 |
| 4 | 3 | 6 | 2 | 7 | 1 | 5 | 8 | 9 |
| 8 | 7 | 1 | 9 | 6 | 5 | 3 | 4 | 2 |
| 9 | 5 | 2 | 4 | 8 | 3 | 7 | 1 | 6 |
| 5 | 1 | 8 | 6 | 2 | 9 | 4 | 7 | 3 |
| 6 | 9 | 3 | 1 | 4 | 7 | 2 | 5 | 8 |
| 2 | 4 | 7 | 5 | 3 | 8 | 6 | 9 | 1 |

Figure 4. An example of a solved Sudoku puzzle - a 9×9 Latin square. The principle of perfect shuffling also holds true in each of the marked 3×3 squares.

| | | | | | | | | |
|---|---|---|---|---|---|---|---|---|
| 9 | 1 | 2 | 3 | 4 | 5 | 6 | 7 | 8 |
| 1 | 2 | 9 | 4 | 5 | 3 | 7 | 8 | 6 |
| 2 | 9 | 1 | 5 | 3 | 4 | 8 | 6 | 7 |
| 3 | 4 | 5 | 6 | 7 | 8 | 9 | 1 | 2 |
| 4 | 5 | 3 | 7 | 8 | 6 | 1 | 2 | 9 |
| 5 | 3 | 4 | 8 | 6 | 7 | 2 | 9 | 1 |
| 6 | 7 | 8 | 9 | 1 | 2 | 3 | 4 | 5 |
| 7 | 8 | 6 | 1 | 2 | 9 | 4 | 5 | 3 |
| 8 | 6 | 7 | 2 | 0 | 1 | 5 | 3 | 4 |

*a*)

| | | | | | | | | |
|---|---|---|---|---|---|---|---|---|
| 9 | 1 | 2 | 3 | 4 | 5 | 6 | 7 | 8 |
| 1 | 2 | 9 | 4 | 5 | 3 | 7 | 8 | 6 |
| 2 | 9 | 1 | 5 | 3 | 4 | 8 | 6 | 7 |
| 3 | 4 | 5 | 6 | 7 | 8 | 9 | 1 | 2 |
| 4 | 5 | 3 | 7 | 8 | 6 | 1 | 2 | 9 |
| 5 | 3 | 4 | 8 | 6 | 7 | 2 | 9 | 1 |
| 6 | 7 | 8 | 9 | 1 | 2 | 3 | 4 | 5 |
| 7 | 8 | 6 | 1 | 2 | 9 | 4 | 5 | 3 |
| 8 | 6 | 7 | 2 | 9 | 1 | 5 | 3 | 4 |

*b*)

Figure 5. A Latin square of order 9 with a chromatic number of 3 (three transversals), which are highlighted in different colors (a), and a covering of the square by these transversals (b). In the general case, the elements of a transversal do not necessarily have to lie on the diagonals of the Latin square.

There are other studies related to the investigation of the structure of Latin squares using group theory, for example, regarding the presence of a group of autoparatopy in their set [23]. We believe that such studies can serve as a mathematical foundation for new ideas in the search for superstructures in high-entropy alloys.

It is likely that a certain hidden symmetry in alloys can also be related, within the framework of the proposed model, to orthogonal Latin squares. Two nth-order Latin squares, $L = (l_{ij})$ and $K = (k_{ij})$, are called orthogonal if all ordered pairs $(l_{ij}, k_{ij})$ are distinct, though the elements within each pair need not be distinct. Using, for example, two orthogonal squares to tile the entire plane (space) in a checkerboard pattern can yield nontrivial high-order symmetries.

It can be assumed that the problem of the chromatic number of a Latin square [26, 27] may also serve as a basis for searching for superstructures in an entropy alloy. The chromatic number of a Latin square is the smallest number of partial transversals that, together, completely cover the square's cells. A transversal of a Latin square of order n is a set of n distinct elements in which no two elements are from the same row and no two elements are from the same column. A specific example is shown in Figure 5.

Finally, let's consider a problem that can be solved within the framework of the proposed model. If we have a two-dimensional structure of the high-entropy alloy type, where N components are perfectly mixed on a plane, what is the probability P(N) of such a structure forming, depending on the number of components? Obviously, we must find the total number of Latin squares L(N) of order N and normalize this by the total number of possible combinations $K_{tot}$ for arranging $N^2$ elements in a square with $N^2$ cells.

First, let's find the total number of arrangements of $N^2$ elements in $N^2$ cells, taking into account that there are N distinct elements (letters), and that permutations of identical letters/numbers between cells are considered the same combination, since identical atoms are indistinguishable. There are a total of $[N^2]!$ permutations of all elements. In the general case of a D-dimensional system, there are $[N^D]!$ such permutations. Identical atoms placed in the cells can be swapped between cells - they are indistinguishable. This results in the same state of the system. The number of such permutations is $(N!)^N$ [28]. Then the total number of possible combinations is $K_{tot}$

$$K_{tot} = \frac{[N^D]!}{(N!)^N} \tag{15}$$

The number of Latin squares is calculated exactly for N ≤ 11: L(2) = 2, L(3) = 12; L(4) = 576; L(5) = 161280; L(6) = 812 851 200; L(7) = 61 479 419 904 000; L(8) = 108 776 032 459 082 956 800 [29]. Then, for the probability of realizing a structure with perfect mixing in an alloy with N components, we obtain the following expression:

$$P(N) = \frac{L(N)(N!)^N}{[N^2]!}. \tag{16}$$

For N=2 - $P(2) = 1/3 = 0.33$; $P(3) \approx 0{,}007142857$; $P(4) \approx 9{,}1 \times 10^{-6}$; $P(5) \approx 2.6 \times 10^{-10}$; $P(6) \approx 3 \times 10^{-16}$; $P(7) \approx 8 \times 10^{-24}$; $P(8) \approx 6 \times 10^{-33}$. This is quite correlated with the estimations of Ref. [30]. The numbers speak for themselves. In a three-dimensional system, these probabilities are expected to be lower. From the point of view of this estimation, even five-component alloys have a vanishingly low probability of realizing the perfect mixing state. For a larger number of components, such an ideal mixing looks unattainable without additional factors for the redistribution of elements in the alloy structure.

To understand the general trend for the probability under consideration, an approximate analytical estimate is of particular interest. Thus, there is an analytical estimate for the upper and lower bounds of the number of Latin squares [19]:

$$\prod_{i=1}^{N}(i!)^{N/i} \geq L(N) \geq \frac{(N!)^{2N}}{N^{N^2}}. \tag{17}$$

However, for small N ~3-7, inequality (17) still gives a rough estimate. In this case, the author dares to offer his analytical estimate of the number of Latin squares for small N. To do this, we use the corner cell of a square as a basis. It can be selected in N ways. Isotopic Latin squares are obtained by permutations of N-1 of the remaining row (their (N-1)! ) and N-1 remaining columns (there are (N-1)! of them too). When any of the N numbers is located in the corner of the square, we exclude from consideration all configurations when the same number is in other cells of the row (another N-1 variants) and in other cells of the column (another N-1 variants). This follows simply from the definition of the Latin square. Then we must multiply the above number of combinations by $(N-1)^2$. However, among such combinations, some will match - they turn out to be a mirror image relative to the diagonals. Let us estimate (without rigorous proof) that such mirror symmetric combinations reduce the number of combinations from N-1 to ~(N-2). In general, instead of N-2, there should be a more complex and precise expression. Then for the number of Latin squares we get the following estimate:

$$L_*(N) \approx N\,(N-2)^2[(N-1)!]^2 \tag{18}$$

Для N от 2 до 4 эта формула дает точное значение числа латинских квадратов. Для больших значений N число квадратов занижено. Так, для N=5 $L_*(5) = 72\,000$, в то время как точное значение 161280. Тем не менее, эта оценка несколько точнее (17) для малых N, которые нас интересуют. Тогда для вероятности $P(N)$ получаем следующую оценку.

For N from 2 to 4, this formula gives the exact value of the number of Latin squares. For large values of N, the number of squares is underestimated. So, for N=5 $L_*(5) = 72\,000$, while the exact value is 161 280. Nevertheless, this estimate is somewhat more accurate (17) for the small N that we are interested in. Then, for the probability P(N), we get the following relation.

$$P(N) \approx \frac{N\,(N-2)^2[(N-1)!]^2(N!)^N}{[N^2]!}. \tag{19}$$

Asymptotically, for large N, the probability P(N) behaves as

$$ln\,P(N) \sim -\,N^2\,ln\left(\frac{N}{e}\right) \tag{20}$$

Probably, these estimates provide an additional reason to weaken the requirement of equiatomicity of the multicomponent composition in the practical consideration of high-entropy alloys. On the other hand, the results of such a consideration make it possible to identify alloys with a very narrow concentration range near the equiatomic one among high-entropy alloys and name them maximum entropy alloys (MEA).

Conclusion.

The paper proposes a concept for a model of the structure of high-entropy alloys based on the mathematics of Latin squares. This model describes the perfect mixing of a large number of components. The features of the model are discussed. The directions of searching for superstructures in high-entropy alloys based on this model are proposed. The probability of perfect mixing of multicomponent alloys was calculated, which turned out to be extremely low.